# Photonic dark matter portal revisited

S A Alavi and F S Kazemian
Department of Physics, Hakim Sabzevari University, Sabzevar, Iran

E-mail: alavi@hsu.ac.ir; alaviag@gmail.com

In our previous paper [1], we studied a model of dark matter (DM) in which the hidden sector interacts with standard model particles via a hidden photonic portal (HP). We investigated the effects of this new interaction on the hydrogen atom and obtained an upper bound for the coupling of the model as $f \leq 10^{-12}$. In this work, we study the effects of HP on two interesting exotic atoms namely muonium and positronium. We obtain a tighter upper limit on the coupling as $f \leq 10^{-13}$. We also calculate the change (shift) in the Aharonov-Bohm phase due to HP and find that the phase shift is negligibly small (for DM particles mass in the GeV range).
Recently a 3.5 keV X-ray line signal observed in the spectrum of 73 galaxy clusters, reported by the XXM-Newton X-ray observatory. Since in HP model the DM particles can decay directly into photons, so we finally calculate the value of the coupling constant f using the condition $\Delta E_{DM} = 3.5 \, keV$.

## 1. Introduction

The nature of dark matter (DM), namely its identity, properties and non-gravitational interactions, remains one of the most important unsolved problems in physics. Attempting to answer the question of how we can find conclusive evidence for dark matter through direct detection experiments is of great importance to physicists.

Various DM models and candidates have been proposed over the years. One possibility is that DM couples to the visible sector through the Higgs portal [2]. Another possibility is coupling through the photonic portal [3]. Coupling through the axion, neutrino and vector portals have also been proposed [4].

## 2 The model

In [5], the author proposed a model of a hidden sector of the universe consisting of sterile spin-1/2 fermions ("sterinos") and sterile spin-0 bosons ("sterons") interacting weakly through the mediation of sterile quanta of an antisymmetric-tensor field $A_{\mu\nu}$ ("A bosons"), but with a more intense coupling than through universal gravity. It is assumed that these sterile particles of the hidden sector can communicate with the standard model (SM) sector through the electromagnetic field $F_{\mu\nu} = \delta_\mu A_\nu - \delta_\nu A_\mu$ weakly coupled to the A-boson field $A_{\mu\nu}$ [5]:

$$L = \frac{1}{2}\sqrt{f}(\varphi F_{\mu\nu} + \xi \bar{\psi}\sigma_{\mu\nu}\psi)A^{\mu\nu}. \tag{1}$$

$\sqrt{f}$ and $\sqrt{f}\xi$ are small coupling constants. It is supposed that $\varphi = <\varphi>_{vac} + \varphi_{ph}$ with a spontaneously nonzero vacuum expectation value $\langle\varphi\rangle_{vac} \neq 0$. This coupling of photons to the hidden sector is called the "photonic portal". The A-boson kinetic and SM electromagnetic Lagrangians together with Lagrangian lead to the following field equations for $F_{\mu\nu}$

and $A_{\mu\nu}$[6]:

$$\partial^V(F_{\mu V} + \sqrt{f}\varphi A_{\mu V}) = -j_\mu \tag{2}$$

$$(\partial_\mu\partial^\mu - \tilde{M}^2)\partial^v A_{\mu v}{}^{(vac)} = \sqrt{f} <\varphi>_{vac} j_\mu \tag{3}$$

where $\tilde{M}^2 = M^2 + f<\varphi>_{vac}$, $j_\mu$ and $M$ are the SM electromagnetic current and the mass scale of the A-boson respectively. The field equation (2) is the modified Maxwell's equations in the presence of the hidden sector interacting with the SM sector through the photonic portal. For a SM point charge at rest at $\vec{x'}$, $j_\mu(\vec{x}) = g_{\mu 0} e' \delta^{(3)}(\vec{x} - \vec{x'})$ and from Eq.(3), we have [7]:

$$\partial^v A_{\mu v}{}^{(vac)}(x) = -g_{\mu 0}\sqrt{f}<\varphi>_{vac} \frac{e'}{4\pi|\vec{x}-\vec{x'}|} e^{-\tilde{M}|\vec{x}-\vec{x'}|} \tag{4}$$

Using Eq. (4), it can be shown that the hidden sector correction $\delta V^{vac}$ to the Coulomb potential $V = \frac{ee'}{4\pi|\vec{x}-\vec{x'}|}$ is given by the following expression [7]:

$$\delta V^{(vac)} = \frac{f\langle\varphi\rangle_{vac}^2}{\tilde{M}^2} \frac{ee'}{4\pi|\vec{x}-\vec{x'}|}\left(1 - e^{-\tilde{M}|\vec{x}-\vec{x'}|}\right) + O(f^2) \tag{5}$$

### 3. The effects of hidden photonic portal on some exotic atoms

3.1. *The Positronium*

The positronium atom provides an excellent test for studing the relativistic bound state two body systems and quantum electrodynamics corrections to that system. Therefore in this section we study the effects of HP on the positronium atom. The Hamiltonian of the system is given by $H = \frac{\vec{p}^2}{2\mu} + V(\vec{r})$, where $\mu = \frac{m_e}{2}$ is the reduced mass, so the energies of the system are $E_n^{pos} = \frac{1}{2}E_n^H$.

In Ref. [7], the author considered a value for the coupling $f$ as 0.0917. It is also proposed that $\langle\varphi\rangle_{vac}^2 \sim (10^{-2}-1)M^2$ and $M \sim (75-770)$ GeV, so $\frac{f\langle\varphi\rangle_{vac}^2}{\tilde{M}^2} \sim 0.00092$, which is sufficiently small to allow us to use perturbation theory to find the energy shift due to the potential $\delta V^{(vac)}$ given by Eq.(5). The energy corrections for the ground and one of the first excited states are as follows:

$$\Delta E_{100} = \frac{f\langle\varphi\rangle_{vac}^2}{\tilde{M}^2}\frac{e}{4\pi}(\frac{1}{a^{pos}} + \frac{4}{a^{pos^3}b^2})$$

$$\Delta E_{200} = \frac{f\langle\varphi\rangle_{vac}^2}{\tilde{M}^2}\frac{e}{4\pi}[\frac{1}{4a^{pos}}\{1 - \frac{2}{b^2} - \frac{3}{b^4 a^{pos^2}} - \frac{4}{a^{pos}b^3}\}]$$

where $b = \tilde{M} + \dfrac{2}{a^{pos}}$ and $a^{Pos}$ is the Bohr radius of the positronium atom, so the energy values (for $\tilde{M} = 100 GeV$) are as follows :

$$\Delta E_{100} = 1.5 \times 10^{-2} \tag{6}$$
$$\Delta E_{200} = 6.8 \times 10^{-3}$$

The accuracy of the energy measurement is $10^{-12}$ eV [8] and the values in Eq.(6) are large enough to be measured experimentally, but so far there is no evidence for these energy corrections in the positronium atom. One can therefore find an upper bound for the coupling $f$ using the fact that the energy corrections must be smaller than the accuracy of energy measurement, giving the result:

$$f \leq 10^{-13}$$

So we obtain a tighter upper limit on the coupling compared the value obtained from hydrogen atom $f \leq 10^{-12}$.

3.2. *The muonium atom*

The reduced mass of the muonium ($\mu^+ e^-$) is given by $\mu = \dfrac{m_e m_\mu}{m_e + m_\mu} = 0.99 m_e$ which has almost the same value as the hydrogen atom. The bohr radius of the two atoms also have almost the same values. Therefore the effects of the HP on the muonium atom is almost the same as hydrogen atom which has been studied in [1].

**4. The effects of HP on the Aharonov-Bohm phase**

In the reference frame in which the source of a four vetcor potential $(\varphi, \vec{A})$ is at rest relative to the hidden sector particle, we have the following corrections on the scalar and vector potentials [1]:

$$\delta\varphi = \dfrac{f\langle\varphi\rangle^2_{vac}}{\tilde{M}^2} \dfrac{e}{4\pi r}\left(1 - e^{-\tilde{M}r}\right)$$
$$\vec{\delta A} = 0$$

The scalar and vector potentials are the components of the electromagnetic four-potential, so they transform according to the Lorentz transformations. Therefore in a reference frame in which the source moves with a constant velocity $\vec{v}$ with respect to the dark matter candidate, we will have the following corrections to the scalar and vector potentials:

$$\delta\varphi' = \gamma\,\delta\varphi$$
$$\vec{\delta A'} = \gamma \vec{v} \delta\varphi$$

The vector potential can be written as follows :

$$A'_\mu = A_\mu + \delta A_\mu$$

Where $\delta A_\mu$ is the Corrections due to the presence of hidden photonic dark particle and is given by:

$$\delta A_\mu = \frac{v_\mu}{\sqrt{1-v^2}} \frac{f\langle\varphi\rangle^2_{vac}}{\tilde{M}^2} \frac{e}{4\pi r'}(1-e^{-\tilde{M}r'})$$

here $r' = ((x-vt)^2\gamma^2 + y^2)^{1/2}$, $\gamma = \frac{1}{\sqrt{1-v^2}}$ and $v$ is the relative velocity of the particle with respect to the hidden dark particle. Therefore the corrections due to the hidden photonic sector on the Aharonov-Bohm phase is given by:

$$\Delta\phi = \frac{e}{h}\oint \delta A . dr \tag{7}$$

Now we consider the configuration in Fig.1, in which the magnetic flux can flow inside an impenetrable cylinder:

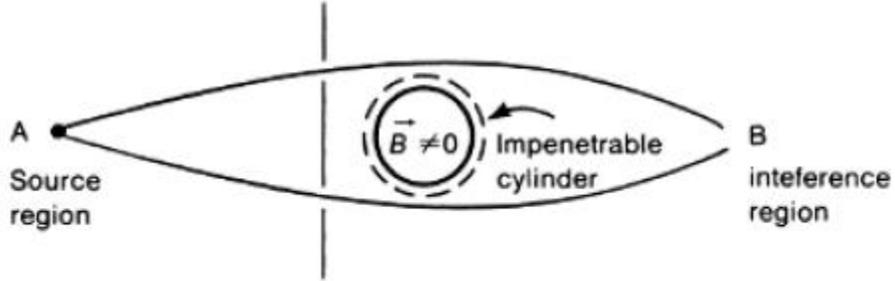

Figure 1: Aharonov bohm effect

In cylindrical coordinates the correction on vector potential can be written as:

$$\vec{\delta A}_\varphi = \frac{-v\,\sin\varphi\,f\langle\varphi\rangle^2_{vac}}{\sqrt{1-v^2}\,\tilde{M}^2} \frac{e}{4\pi r'}\left(1-e^{-\tilde{M}r'}\right)\hat{\varphi} \tag{8}$$

From Eqs.(7),(8), one can calculate $\Delta\varphi$. We found that the corrections due to hidden photonic portal on Aharonov-Bohm phase for different velocities are of the order of $10^{-22}$
(for $M = 100$ GeV).

### 5. Photonic hidden sector and 3.5 keV line from galaxy clusters

Recently a 3.5 keV X-ray line signal has been observed in the spectrum of 73 galaxy clusters which reported by the XXM-Newton X-ray observatory[9]. Many attempts have been made to explain the origin of this keV line. there has been also considerable interest in trying to explain it in terms of dark matter scattering or decays (see [10] and references therein).

In the hidden photonic model discussed in section 2, the A bosons can decay into steron and photon $A \to \varphi \gamma$. The decay rate is given by the following expression [11]:

$$\Gamma = \frac{f}{96\pi} M (1 - \frac{m_\varphi^2}{M^2})^3$$

where $M$, $m_\varphi$ are the mass of the $A$ and $\varphi$ bosons respectively. Assuming $M = 100 GeV$, $m_A - m_\varphi = 3.5 keV$ and using the relation between the life time and the mass of the dark matter [12]:

$$\tau_{DM} = 10^{28} - 10^{29} \sec \frac{7 keV}{m_{DM}}$$

we obtain the following value for the coupling of the model:

$$f = 10^{-25}$$

which is too small to lead to observable effects at collider experiments.

**Conclusions**

We studied the effects of the hidden photonic portal on two exotic atoms i.e., positronium and muonium. We obtained a tighter upper limit for the coupling of the model than the one obtained from hydrogen atom. We also studied the effects of HP on Aharonov-Bohm phase. We found that for DM mass scale 100 GeV, the phase shift (correction) is of the order of $10^{-22}$, which is negligibly small. To explain the recently observed 3.5 keV line signal, we used the condition $\Delta E_{DM} = 3.5\ keV$ and obtained the value $f = 10^{-25}$ for the coupling of the model which is too small to lead observable effects at collider experiments.